\pdfoutput=1

\documentclass[final]{aa}

\usepackage{graphicx}
\usepackage{amssymb,amsmath}
\usepackage[varg]{txfonts}
\usepackage{natbib}
\usepackage{hyperref}
\hypersetup{pdftitle = The title of my PDF, pdfauthor = My name, pdfsubject= The subject, pdfkeywords = keyword1 keyword2 keyword3}
\hypersetup{colorlinks = true, linkcolor = blue, anchorcolor = red, citecolor = blue, filecolor = red, pagecolor = red, urlcolor = blue}

\begin{document}

\title{The grain size survival threshold in one-planet post-main-sequence exoplanetary systems}

\titlerunning{Grain size survival in giant branch systems}
\authorrunning{Zotos \& Veras}

\author{Euaggelos E.~Zotos\inst{\ref{inst1}} \and Dimitri Veras\thanks{STFC Ernest Rutherford Fellow}\inst{\ref{inst2},\ref{inst3}}}

\institute{
Department of Physics, School of Science, Aristotle University of Thessaloniki, 541 24, Thessaloniki, Greece \\
\email{evzotos@physics.auth.gr}\label{inst1}
\and
Centre for Exoplanets and Habitability, University of Warwick, Coventry, CV4 7AL, UK\label{inst2}
\and
Department of Physics, University of Warwick, Coventry, CV4 7AL, UK \\
\email{d.veras@warwick.ac.uk}\\
\label{inst3}
}


\abstract{
The size distribution and orbital architecture of dust, grains, boulders, asteroids, and major planets during the giant branch phases of evolution dictate the preponderance and observability of the eventual debris, which have been found to surround white dwarfs and pollute their atmospheres with metals. Here, we utilize the photogravitational planar restricted three-body problem in one-planet giant branch systems in order to characterize the orbits of grains as the parent star luminosity and mass undergo drastic changes. We perform a detailed dynamical analysis of the character of grain orbits (collisional, escape, or bounded) as a function of location and energy throughout giant branch evolution. We find that for stars with main-sequence masses of $2.0M_{\odot}$, giant branch evolution, combined with the presence of a planet, ubiquitously triggers escape in grains smaller than about 1 mm, while leaving grains larger than about 5 cm bound to the star. This result is applicable for systems with either a terrestrial or giant planet, is largely independent of the location of the planet, and helps establish a radiative size threshold for escape of small particles in giant branch planetary systems.
}


\maketitle

\section{Introduction}
\label{intro}

Growing observations of post-main-sequence planetary systems motivate increasing our understanding of how objects of all sizes from grains to major planets evolve along with their parent star. The over 100 major planets discovered around subgiant or red giant branch stars \citep{refetal2015,gheetal2018,gruetal2018,gruetal2019}\footnote{Sabine Reffert maintains a dedicated database for these planets at http://www.lsw.uni-heidelberg.de/users/sreffert/giantplanets.html.} will continue to increase due to the successful launch of the {\it TESS} mission. Further, at least four debris disks have been detected around subgiant stars, three of which are known to host exoplanets \citep{bonetal2013,bonetal2014}. A tentative detection of an exoplanet has even been reported orbiting an asymptotic giant branch star \citep{keretal2016}.

These giant star planetary systems provide a foundation for understanding their fate after the parent stars have transformed into white dwarfs. Observations of white dwarf planetary systems have yielded high-profile discoveries of major \citep{ganetal2019} and minor \citep{vanetal2015,manetal2019,vanetal2019} planets as well as startling detail in the data, which showcase the importance of understanding the size distribution of particles. Over 40 debris disks have been discovered orbiting white dwarfs \citep{zucbec1987,ganetal2006,farihi2016,denetal2018,manetal2020} and they appear to be almost ubiquitously variable over yearly or decadal timescales \citep{faretal2018,xuetal2018,swaetal2019a,wanetal2019}.

This variability may be explained by dynamical activity. One potential source of activity is the interaction between gas and dust within the disk \citep{bocraf2011,rafikov2011a,rafikov2011b,metetal2012,rafgar2012,kenbro2017b,vanetal2018}, which is notably different from dust-only evolution \citep{kenbro2017a} and gas-only evolution \citep{mirraf2018}. Regardless, the specific abundances and sizes of the particles, which may or may not be involved in the disk evolution, play a crucial role in producing observable signatures.

Although white dwarf disk particles are primarily generated from tidally-induced formation of the disk \citep{graetal1990,jura2003,debetal2012,veretal2014a,makver2019,malper2020a,malper2020b}, contributions from extant submeter-sized particles would arise from radiative drag due to the white dwarf luminosity \citep{veretal2015a,veretal2015b,veras2020} and injection from outer system reservoirs \citep{griver2019,verkur2020}. Hence, the variety and abundance of grains and boulders, which survive giant star evolution \citep{donetal2010,veretal2015c}, represent a crucial component of the planetary system architecture surrounding white dwarfs\footnote{In addition to the dynamical motivation, there is also a strong compositional link between grains, boulders, and minor planets with the material which is eventually accreted onto the white dwarf photosphere \citep[see e.g.,][]{zucetal2007,gaeetal2012,juryou2014,xuetal2017,haretal2018,holetal2018,doyetal2019,swaetal2019b,bonetal2020}.}.

Theoretical studies of major planets orbiting giant branch stars are increasing (for a review, see \citealt*{veras2016}). However, investigations which have considered the interaction between major planets and smaller bodies during this stellar phase have generally treated the smaller bodies as point particles. Few investigations have studied the combined effect of giant branch stellar radiation, mass loss and planetary perturbations on finite-sized small bodies \citep{donetal2010,veretal2019}. Here we also analyze this situation, but from a different perspective: through the photogravitational planar restricted three-body problem (PPRTBP). As an analytic construct, the PPRTBP allows for an extensive exploration of parameter space and quantification of stability at different levels of stochasticity.

The wide sweep of parameter space we perform here yields bounds on the grain size for which escape from the stellar system is predominant. Escape from giant branch systems due to stellar mass loss alone can occur only for objects which are at least about $10^3$~au distant from the parent star \citep{veretal2011,vertou2012,adaetal2013,veretal2014b}. Radiative escape can occur elsewhere for sufficiently small objects. One potential generator of radiative escape is the Yarkovsky effect \citep{veretal2019}, but that force only applies to objects larger than about 0.1 m \citep{veretal2015c,veras2020}. For submeter sized objects, Poynting-Robertson drag creates an inward drift. Hence, such small objects can be ejected only by the presence of a planet.

Here we illustrate how the planet, in combination with the varying luminosity and mass of the star, sweeps away particles as a function of particle size (a similar pursuit in the context of the solar system and with multiple planets and a constant $1L_{\odot}$ luminosity was performed by \citealt*{mormal2002}). We also explore and quantify the dynamics of non-escape orbits in giant branch systems. We do not make any assumptions about the origins of the particles, but note that they can be generated at any time during the giant branch phases due to breakup of asteroids from the YORP effect \citep{veretal2014c,versch2020}.

In Section \ref{TBP}, we describe the PPRTBP in more detail and how it can be applied to giant branch planetary systems. We present our numerical method and our results in Section \ref{resu}, discuss these results in Section \ref{disc} and conclude in Section \ref{conc}.

\section{The Photogravitational Planar Restricted Three-Body Problem}
\label{TBP}

\subsection{Theoretical background}
\label{back}

The Photogravitational Planar Restricted Three-Body Problem (PPRTBP) is equivalent to the Planar Restricted Three-Body Problem, but with a radiating primary. In this setup, massive primary and secondaries orbit one another, and gravitationally perturb a particle that resides in the same plane. The particle is assumed to have a mass small enough to not alter the orbit of the primary and secondary, but large enough to have a finite radius, and be subject to both radiative and gravitational perturbations from the primary. Over the years, the PPRTBP has been investigated in a large series of papers \citep[see e.g.,][]{SMB85,PKM02,P03,P06,KPR06,PKD06,KPP08,MAB09}.

Here, the primary is the star with mass $M$, the secondary is a major planet with mass $M_{\rm plan}$, and the particle is a grain. As is customary with treatments of the three-body problem, we establish the equations in a scale-free manner. We define $m_1 = 1 - \mu$ and $m_2 = \mu$, where
\begin{equation}
\mu \equiv \frac{M_{\rm plan}}{M + M_{\rm plan}}
.
\end{equation}
The star is thought to be located at $(x_1, 0)$, and the planet at $(x_2, 0)$, with $x_1 = -\mu$ and $x_2 = 1 - \mu$. Consequently, the formula of the effective potential of the system reads \citep{murder1999,zotos2015}
\begin{equation}
\Phi(x,y) =  \frac{\left(1 - \beta\right) m_1}{\sqrt{\left(x - x_1\right)^2 + y^2}} + \frac{m_2}{\sqrt{\left(x - x_2\right)^2 + y^2}} + \frac{1}{2}\left(x^2  + y^2 \right),
\label{pot}
\end{equation}
where $\beta$ is the radiation pressure factor, which is be defined and explained in the next subsection.

In a dimensionless corotating frame of reference, the scaled equations for the planar motion of the test particle are
\begin{equation}
\ddot{x} =  2 \dot{y} + \frac{\partial \Phi}{\partial x},
\label{eqmotx}
\end{equation}

\begin{equation}
\ddot{y} = -2 \dot{x} + \frac{\partial \Phi}{\partial y}.
\label{eqmoty}
\end{equation}

Further, for the PPRTBP, there exists one integral of motion, and it is given by the following Hamiltonian function
\begin{equation}
H(x,y,\dot{x},\dot{y}) = - \left(\dot{x}^2 + \dot{y}^2 \right) + 2\Phi(x,y) = C,
\label{ham}
\end{equation}
where $C$ is the conserved Jacobi constant.

\subsection{Application to giant branch planetary systems}
\label{appl}

The key tenet of this work is that the PPRTBP can be applied to giant branch planetary systems. In order to achieve the link, we must consider how the variables listed in the last subsection convert into physical units. First, however, we describe the evolution of a typical giant branch star.

A star which has left the main sequence undergoes significant physical changes on short timescales. These changes include mass loss, radius expansion and luminosity enhancement. The rate of these changes is a strong function of both the star's initial mass and metallicity. Here we consider only Solar-metallicity stars with initial main-sequence masses of $2M_{\odot}$. The reason for this choice is because that mass corresponds to the progenitor mass of most currently observed white dwarfs \citep{treetal2016}, and so is widespread for post-main-sequence planetary systems.

We used the {\tt SSE} code \citep{huretal2000} to model the evolution of these stars. Because we do not model a specific known star, we are not concerned with small differences in the evolution profile that would arise by using a different stellar evolution code. The main-sequence phase of a $2M_{\odot}$ star lasts for approximately 1.2 Gyr. However, the two subphases of interest are the red giant branch (RGB) and asymptotic giant branch (AGB) subphases; for $2M_{\odot}$ stars, the starkest changes occur during the AGB, and in particular along the thermally pulsing AGB (TPAGB) phase.

Overall, during the 322 Myr-long giant branch phases (and primarily along the 1.6 Myr-long TPAGB), $2M_{\odot}$ stars lose 68.2\% of their mass, increase their luminosity up to $9190L_{\odot}$ and increase their radius up to about $392R_{\odot} \approx 1.82$ au\footnote{For perspective, even without the presence of a planet, a mm-sized grain within about 100 au of star at the beginning of the giant branch phases would not survive until the white dwarf phase due to this enhanced stellar luminosity.}. These changes occur in non-steady fashions that are not well-fit by simple analytical functions. For example, between the tip of the RGB and the start of the TPAGB phase, the star traverses the horizontal branch and early AGB branch, where the luminosity first experiences a slight dip followed by a steady increase.

Usually, the PPRTBP is applied with fixed masses, radii and luminosity. Here, all three variables are functions of time, requiring us to take snapshots at particular epochs of interest.  Nevertheless, the time dependence in this work does not change the traditional definition of the radiation-based variable $\beta$ from Eq. (\ref{pot}):

\begin{equation}
\beta \equiv \frac{F_{\rm rad}}{F_{\rm gra}}.
\end{equation}

\noindent{}Here, $F_{\rm rad}$ and $F_{\rm gra}$ refer, respectively, to the forces from stellar radiation and gravity. Expressions for both are provided in \citep{veretal2015c}. The gravitational force is the sum of the unperturbed two-body force as well as the force applied due to stellar mass loss:

\begin{equation}
F_{\rm gra} \approx -\frac{Gm_{\rm p}M\vec{r}}{r^3}
\end{equation}

\noindent{}where $m_{\rm p}$ is the grain mass (which is small enough to not perturb the planet-star orbit). The radiative force is the sum of several terms

\begin{equation}
F_{\rm rad} \approx -\frac{A L}{4 \pi c r^2}
\bigg[Q_{\rm abs} \vec{I} + Q_{\rm ref} \vec{I} + k \left(Q_{\rm abs} - Q_{\rm ref}\right) \vec{Y}  \bigg] \vec{\iota}
\label{Frad}
\end{equation}

\noindent{}with \citep{buretal1979,buretal2014}

\begin{equation}
\vec{\iota} = \left(1 - \frac{\vec{v} \cdot \vec{r}}{cr}  \right)
\frac{\vec{r}}{r}
-
\frac{\vec{v}}{c}.
\end{equation}

The variables in Eq. \ref{Frad} include the grain's displacement $\vec{r}$ and velocity $\vec{v}$ relative to the stellar center, the speed of light $c$, and the grain's cross-sectional area $A$. The constants $Q_{\rm abs}$ and $Q_{\rm ref}$ respectively refer to the grain's absorption efficiency and reflecting efficiency (or albedo). $k$ is a constant, $\vec{I}$ is the 3x3 unit matrix and $\vec{Y}$ is the 3x3 Yarkovsky matrix (given explicitly in \citealt*{veretal2015c}). Note that the radiative force is inversely proportional to the square of the displacement.

In order to better elucidate the dependence of $\beta$ on the stellar parameters, consider a snapshot in time where both $\vec{r}$ and $\vec{v}$ are fixed. Also assume that the grain properties remains fixed (an assumption that we make throughout our simulations). Then

\begin{equation}
\beta \propto \frac{L}{M}
.
\label{eqLM}
\end{equation}

Because we aim for an approximate treatment here (as we are not analyzing a particular star), we can simplify the expression for $q$. First, the maximum value of the absolute value of the expression in the square brackets is 2, and we apply this bound here. Further, because $\left|\vec{\iota} \right| \approx 1$, we ignore this vector. We also assume that the grain is spherical such that $A=\pi R_{\rm p}^2$, where $R_{\rm p}$ is the radius of the grain. This variable represents the key one for which the results in this paper are calibrated.

\section{Simulations}
\label{resu}

Having described the underlying physics, we now proceed with our numerical simulations. We first describe our numerical method and object properties in detail, before reporting the results.

\subsection{Numerical method}
\label{numres}

Our goal is to describe the character of a grain's orbit at given snapshots in time for particular locations of the grain. We do so by performing a succession of relatively short ($4 \times 10^4 - 4 \times 10^5$ yr) numerical integrations in $(x,y)$ space. We then, following the approach introduced in \citet{N04,N05}, classify the orbit into one of the following categories, with the corresponding colors on many of the figures in this paper: (i) Bounded, regular motion (green), (ii) Bounded, sticky motion (purple), (iii) Bounded, chaotic motion (yellow), (iv) Collision with the star (dark blue), (v) Collision with the exoplanet (red), and (vi) Escape (cyan).

We would like to clarify that sticky orbits are those trajectories which evolve as regular ones for large time intervals, before they finally reveal their true chaotic nature. Usually, the respective starting conditions of sticky orbits appear as lonely, isolated points which are deeply hidden inside the chaotic regions and also in the near vicinity of the stability islands.

In order to classify the different types of bounded motion, we utilize the Smaller Alignment Index (SALI) \citep{S01} method. Our collision detection is triggered when the distance between the grain and the star or planet becomes equal with the radius of the larger body. Our escape detection is triggered when the total inertial energy $E$ of the grain becomes positive, where

\begin{equation}
E = \frac{1}{2} \left(\upsilon_x^2 + \upsilon_y^2 \right) - \Phi(x,y) + \frac{1}{2}\left(x^2  + y^2 \right),
\label{iner}
\end{equation}
such that
\begin{equation}
\upsilon_x = \dot{x} - y, \ \ \ \upsilon_y = \dot{y} + x.
\label{vel}
\end{equation}
The condition $E > 0$ occurs when the grain speed exceeds the local escape speed. Needless to say, the total inertial energy does not remain constant during the numerical integration.

In every case, we achieve high-resolution coverage of phase space by integrating $1024 \times 1024$ starting conditions on different types of two-dimensional maps. Further, in each case, the grain is launched from the horizontal $x$ axis at $x = x_0$, $y_0 = \dot{x_0} = 0$. The initial value of $\dot{y}$ is derived from the Jacobi integral of motion (see Eq. \ref{ham}).

The equations of motion (\ref{eqmotx}-\ref{eqmoty}), along with the set of the variational equations, were numerically integrated by using a variable time step Bulirsch-Stoer integrator, while the corresponding routine was written in \verb!FORTRAN 77! \citep{PTVF92}. Throughout the computations, the Jacobi constant $C$ was conserved to the order of $10^{-14}$.

\begin{figure*}[!ht]
\centering
\resizebox{\hsize}{!}{\includegraphics{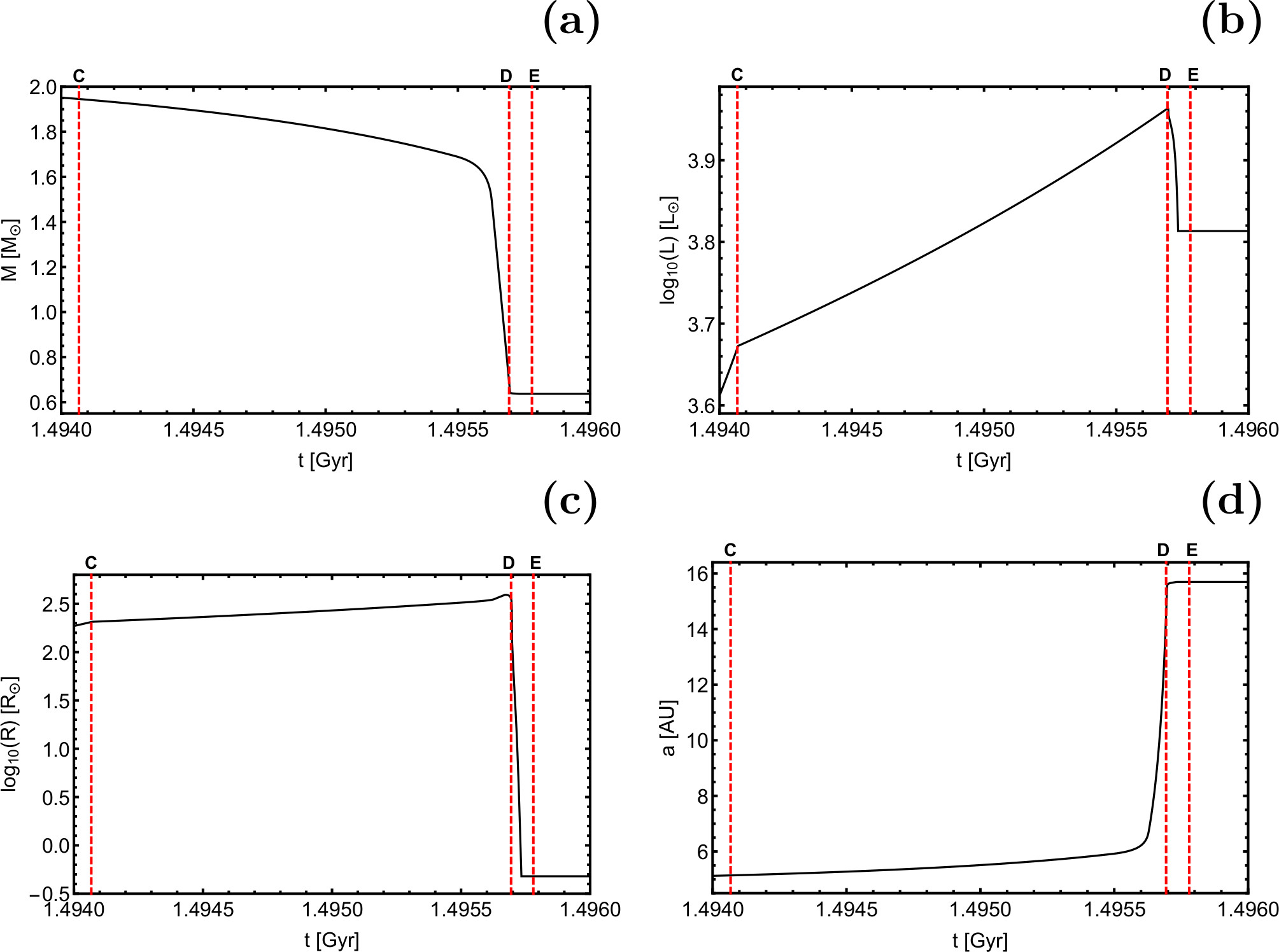}}
\caption{Time-evolution of the mass, luminosity and radius of the star (top three panels) and the semi-major axis of the planet-star system (bottommost panel). The red, vertical, dashed lines correspond to the critical stages of the stellar evolution, which are (``C'': Start of TPAGB; ``D'': Tip of AGB; ``E'': Start of WD).}
\label{evols}
\end{figure*}

The total time of the numerical integration was set to $5000$ time units, so as to ensure that all trajectories have enough time to unveil their true character. One time unit corresponds to the planet-star orbital period. As computed from Table \ref{evols}, this time period can vary from about 7.9 yr to 78 yr depending on the starting time. Hence, the total integration time for each individual simulation is in the approximate range of $4 \times 10^4 - 4 \times 10^5$ yr. These integrations correspond to a specific fixed time frame of the stellar evolution. Consequently, the results are approximate: the accuracy of the final result depends both on the assumption of a fixed stellar evolutionary time-frame and the convergence of the Bulirsch-Stoer integrator as a function of time.

The average required computation time, per color map, varied between 0.5 hours and 4.5 days by using a Quad-Core i7 4.0 GHz CPU (without using an MPI code) with 32 GB of RAM. In order to create the graphics in this paper, we used the latest version 12.0 of Mathematica$^{\circledR}$ \citep{W03}.

\subsection{Object properties}
\label{objprop}

\begin{figure}[!ht]
\centering
\resizebox{\hsize}{!}{\includegraphics{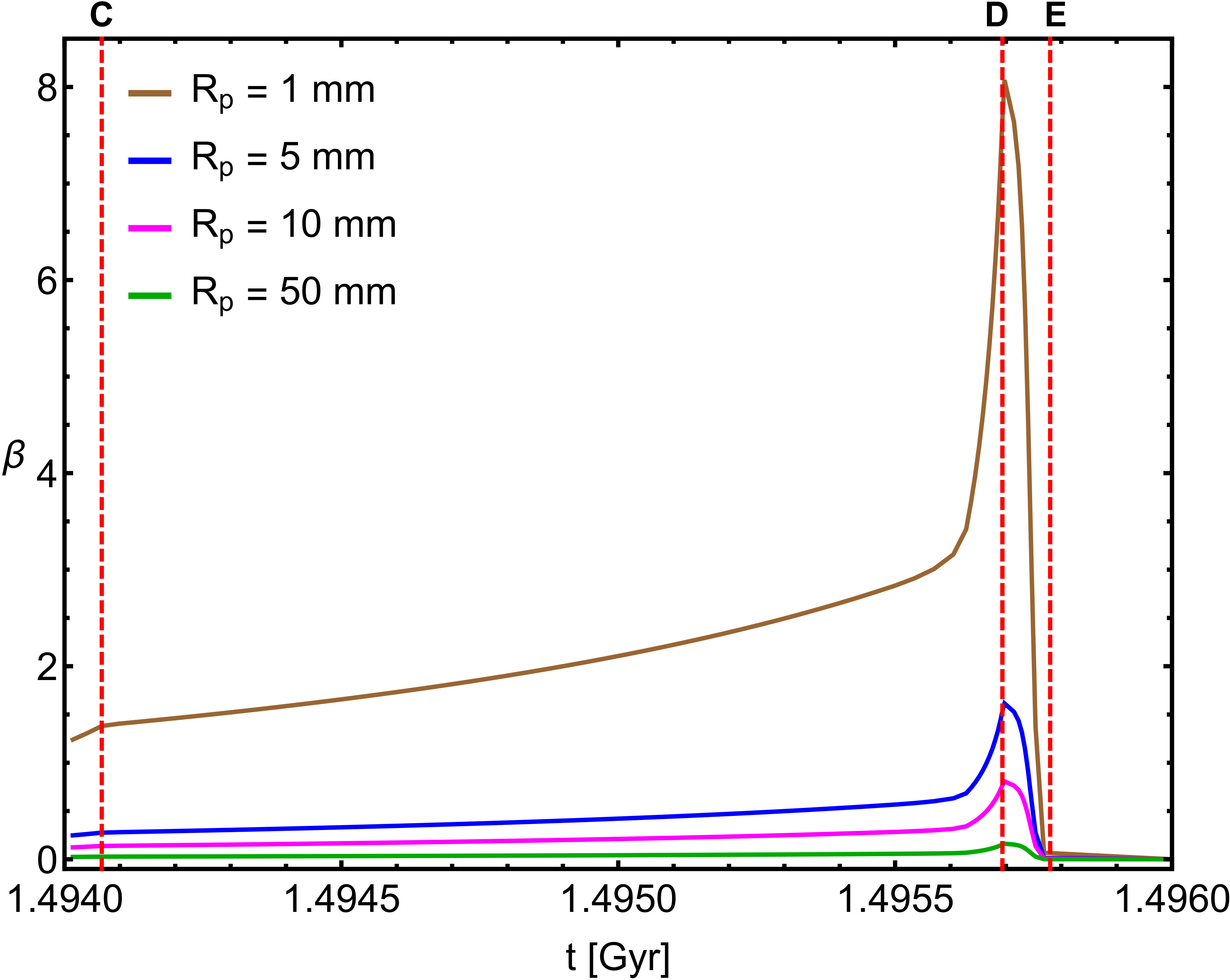}}
\caption{Time-evolution of the radiation pressure factor $\beta$ for grains of four different radii. The red, vertical, dashed lines are the same as those in Fig.~\ref{evols}. Although not discernable on the plot, the Tip of the AGB does not coincide exactly with the peak of $\beta$ because at that time the stellar mass is still decreasing. However, the time between both peaks is just about 1600 yr.}
\label{evolr}
\end{figure}

\subsubsection{Stellar evolution properties}

Understanding how the star evolves across the RGB and AGB subphases is crucial. Hence, we report important values throughout stellar evolution in Table \ref{tab1}, and in Fig.~\ref{evols} we present the time evolution of the key properties of the star (its mass $M$, luminosity $L$ and radius $R$) in the top three panels. We focus on the most relevant time interval, which is the entire AGB (recall also that the RGB epoch provides a negligible change in stellar parameters compared to the AGB). Labelled dashed vertical red lines mark important events in this star's evolution.

The bottom panel of Fig.~\ref{evols} displays a representative semimajor axis ($a$) evolution of the mutual orbit between the star and planet. The PPRTBP formalism is general enough so that actually any initial semimajor axis may be chosen. For increased clarity, throughout the paper we have chosen 5 AU as the initial semimajor axis. This choice is not arbitrary: closer-in planets may be engulfed by giant branch tides \citep{kunetal2011,musvil2012,adablo2013,norspi2013,viletal2014,madetal2016,staetal2016,galetal2017,raoetal2018,sunetal2018}, whereas as planets much further away are less likely due to formation constraints.

For an initial separation of 5 AU, the semimajor axis evolution in Fig.~\ref{evols} is inversely proportional to the mass which remains in the star, based on the two-body problem with variable mass \citep{omarov1962,hadjidemetriou1963}. Only for distances beyond about $10^3$ AU, when the system is said to no longer be ``adiabatic'', does this proportionality break down \citep{veretal2011}. Hence, the bottom panel is actually redundant, being determined simply from the mass evolution in the top panel. Importantly, the eccentricity of the orbit does not change appreciably in this adiabatic regime. Hence, here we assume that the initial eccentricity is zero and does not change throughout the evolution.

\begin{table*}[!t]
  \caption{Parameters for the most important stages of the stellar evolution. The last column gives the radius of the grain corresponding to $\beta = 1$.}
  \centering
  \label{tab1}
  \setlength{\tabcolsep}{3pt}
  \begin{tabular}{@{}lcccccc}
    \hline
    Stage & $t [{\rm Gyr}]$ & $M [M_{\odot}]$ & $R [R_{\odot}]$ & $L [L_{\odot}]$ & $a [{\rm AU}]$ & $R_{\beta=1}[{\rm mm}]$ \\
    \hline
    S: Main sequence            & 0.00000 & 2.0000 & 1.6   & 16   & 5.0000 & 0.0046 \\
    A: Start of RGB             & 1.17357 & 1.9999 & 5.8   & 20   & 5.0003 & 0.0058 \\
    B: Tip of RGB               & 1.19647 & 1.9979 & 28    & 240  & 5.0052 & 0.070  \\
    C: Start of TPAGB           & 1.49407 & 1.9452 & 210   & 4700 & 5.1409 & 1.4    \\
    D: Tip of AGB               & 1.49569 & 0.6613 & 320   & 9200 & 15.122 & 8.0    \\
    E: Start of WD              & 1.49578 & 0.6365 & 0.012 & 23   & 15.710 & 0.021  \\
    \hline
  \end{tabular}
\end{table*}

\begin{figure*}[!t]
\centering
\resizebox{0.7\hsize}{!}{\includegraphics{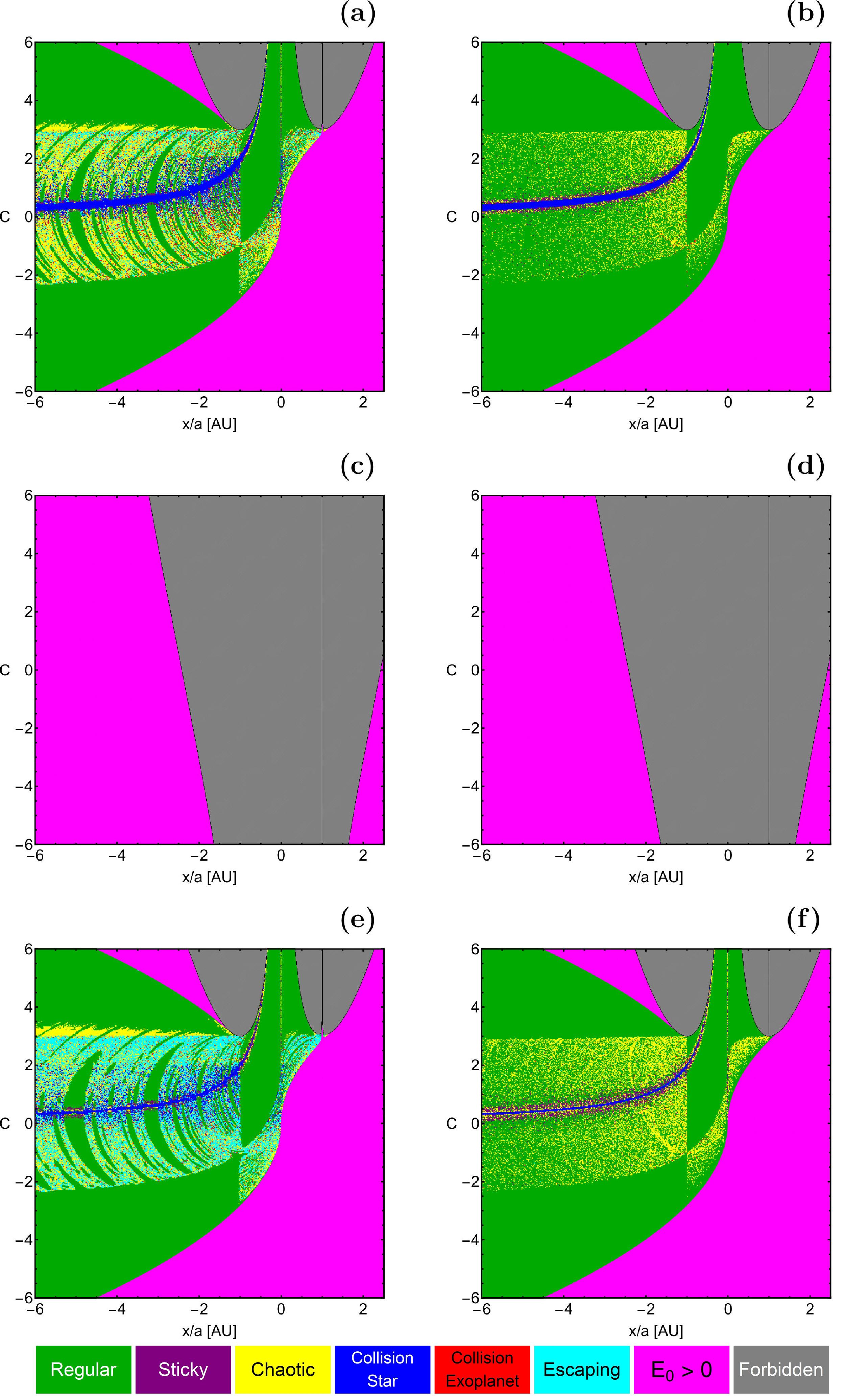}}
\caption{Orbit types for $R_{\rm p} = 1$ mm grains as a function of the coordinate $x$ and Jacobi constant $C$. The left and right panels correspond respectively to Jupiter-like and Earth-like exoplanets, and the top, middle and bottom rows respectively correspond to the main-sequence stage (with $\beta = 0.00058$), tip of the AGB (with $\beta = 8.11062$), and WD stage (with $\beta = 0$).}
\label{xc01}
\end{figure*}

\begin{figure*}[!t]
\centering
\resizebox{\hsize}{!}{\includegraphics{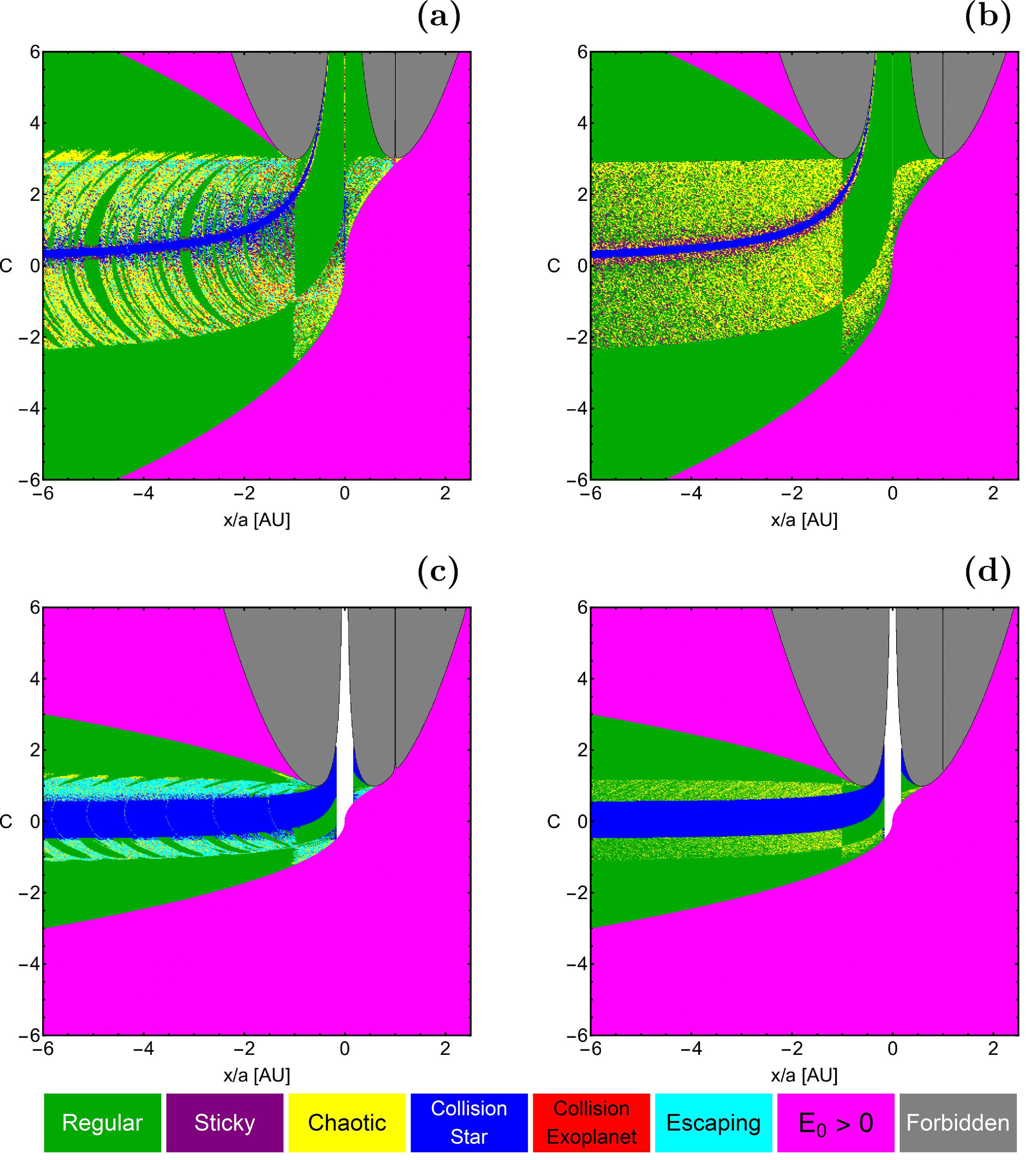}}
\caption{Same as Fig.~\ref{xc01}, but for $R_{\rm p} = 10$ mm and without the WD results. The main-sequence stage (top row) corresponds to $\beta = 0.00058$ and the tip of the AGB stage (bottom row) corresponds to $\beta = 0.811062$.}
\label{xc10}
\end{figure*}

\begin{figure*}[!t]
\centering
\resizebox{\hsize}{!}{\includegraphics{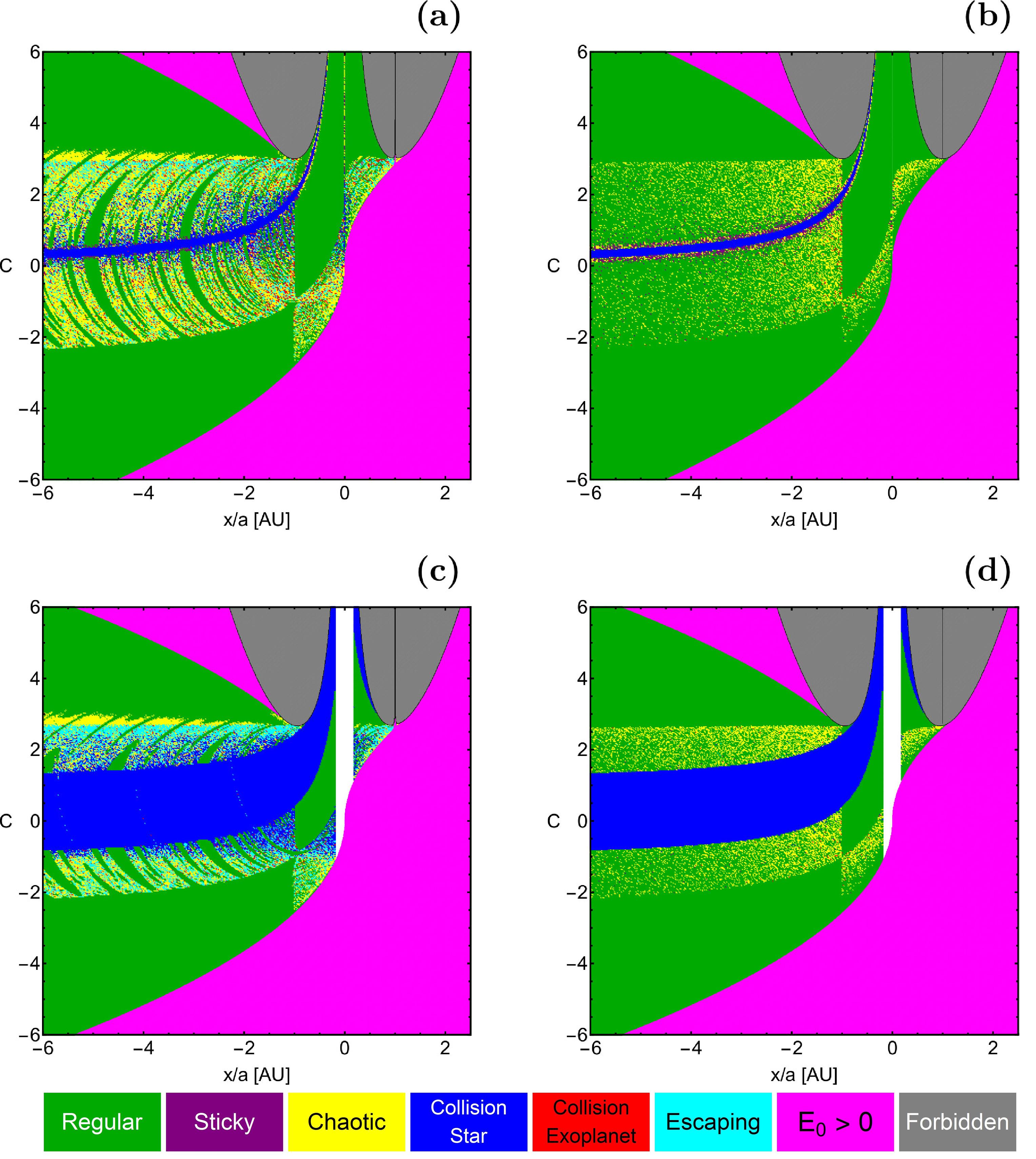}}
\caption{Same as Fig.~\ref{xc10}, but for $R_{\rm p} = 50$ mm.  The main-sequence stage (top row) corresponds to $\beta = 0.00016$ and the AGB tip stage (bottom row) corresponds to $\beta = 0.162212$.}
\label{xc50}
\end{figure*}

\begin{figure*}[!t]
\centering
\resizebox{\hsize}{!}{\includegraphics{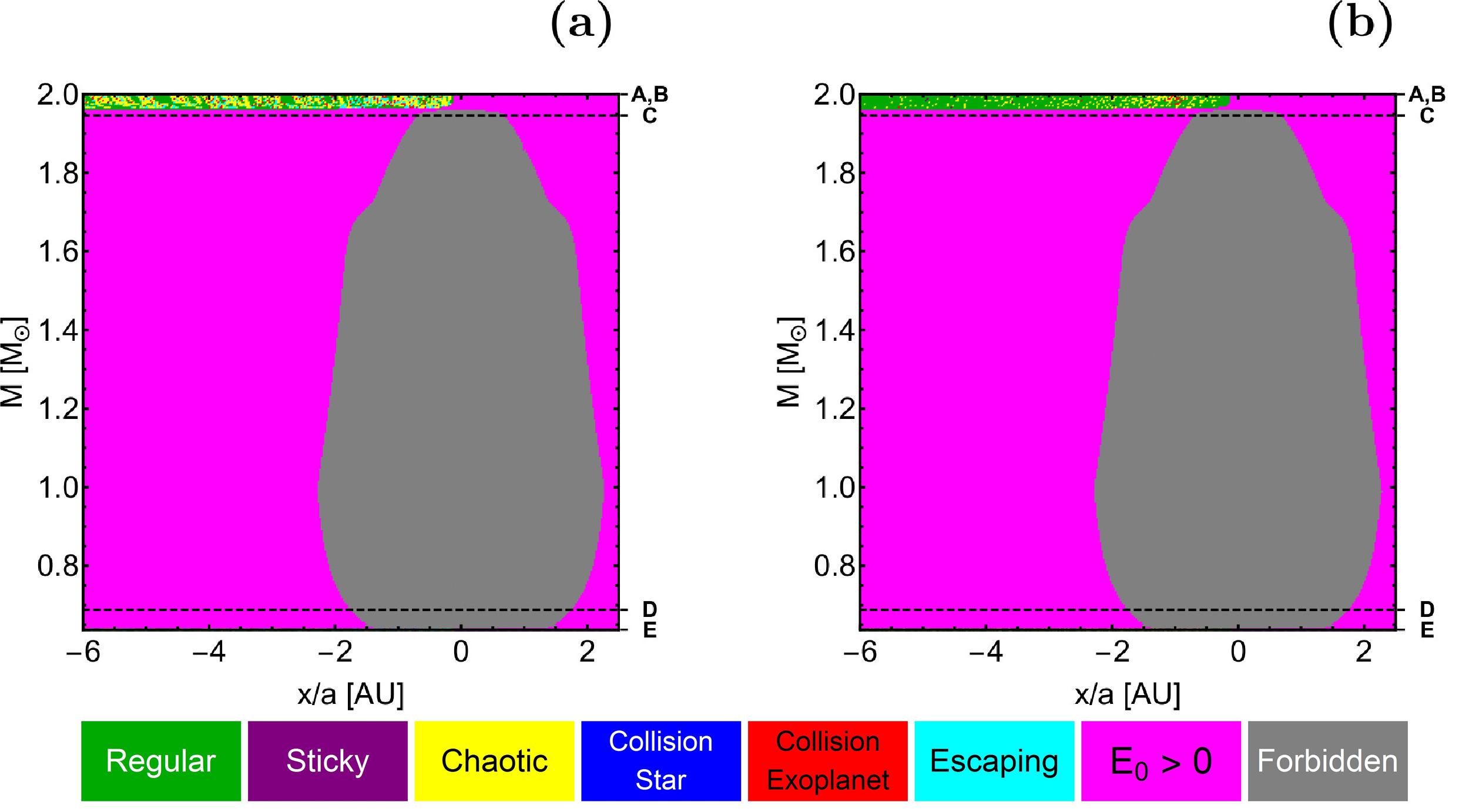}}
\caption{Orbit types for $R_{\rm p} = 1$ mm grains as a function of the coordinate $x$ and the mass of the star $M$. The left and right panels correspond respectively to Jupiter-like and Earth-like exoplanets, and in both cases $C = -1$. The horizontal dashed lines correspond to (``A'' = Start of RGB, ``B'' = Tip of RGB, ``C'' = Start of TPAGB,  ``D'' = Tip of AGB, ``E'' = Start of WD).}
\label{xm01}
\end{figure*}

\begin{figure*}[!t]
\centering
\resizebox{0.7\hsize}{!}{\includegraphics{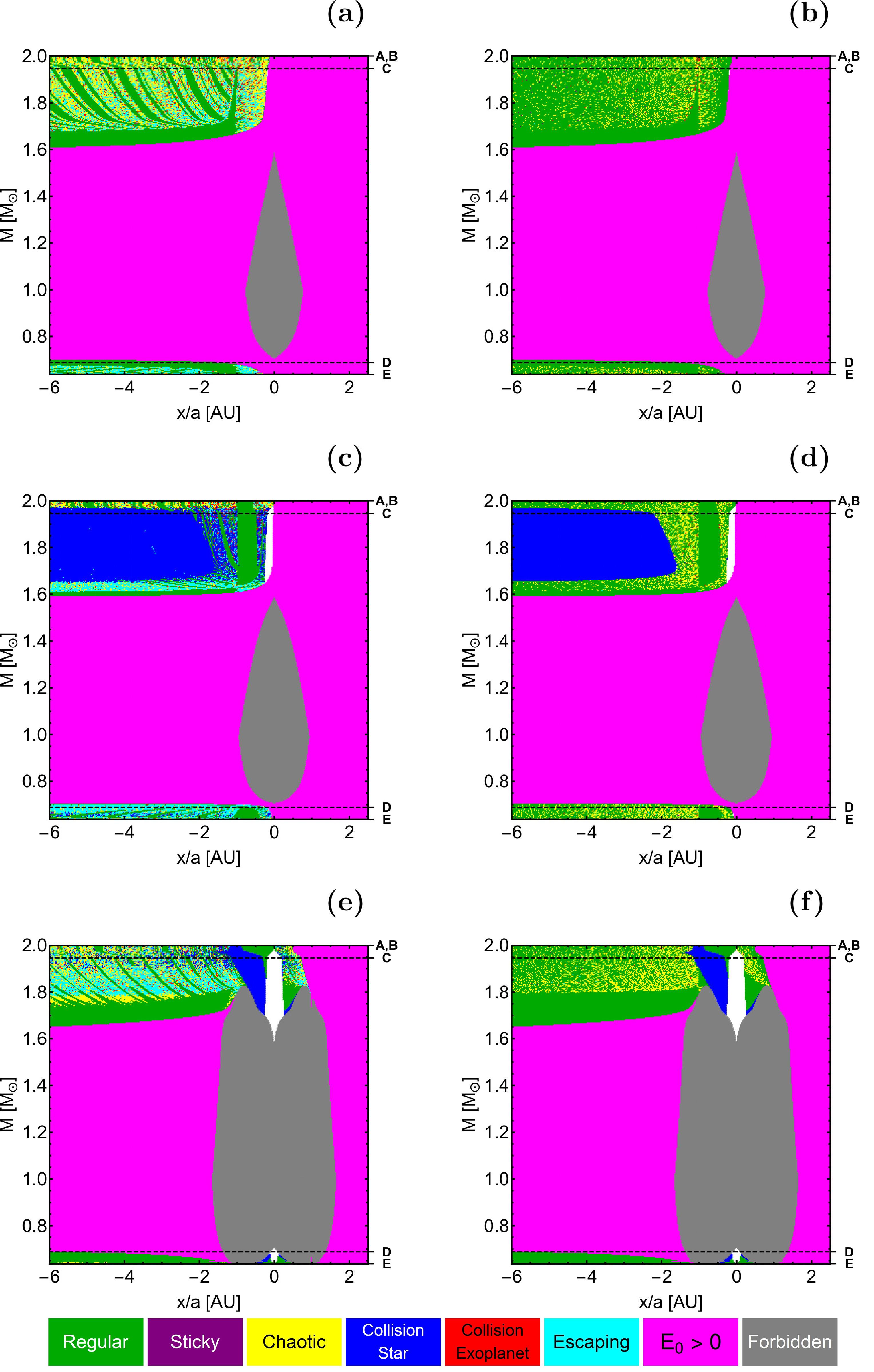}}
\caption{Same as Fig.~\ref{xm01}, but for $R_{\rm p} = 5$ mm. The upper, middle and lower rows correspond respectively to $C=-1, -0.4$ and $-2$.}
\label{xm05}
\end{figure*}

\begin{figure*}[!t]
\centering
\resizebox{0.7\hsize}{!}{\includegraphics{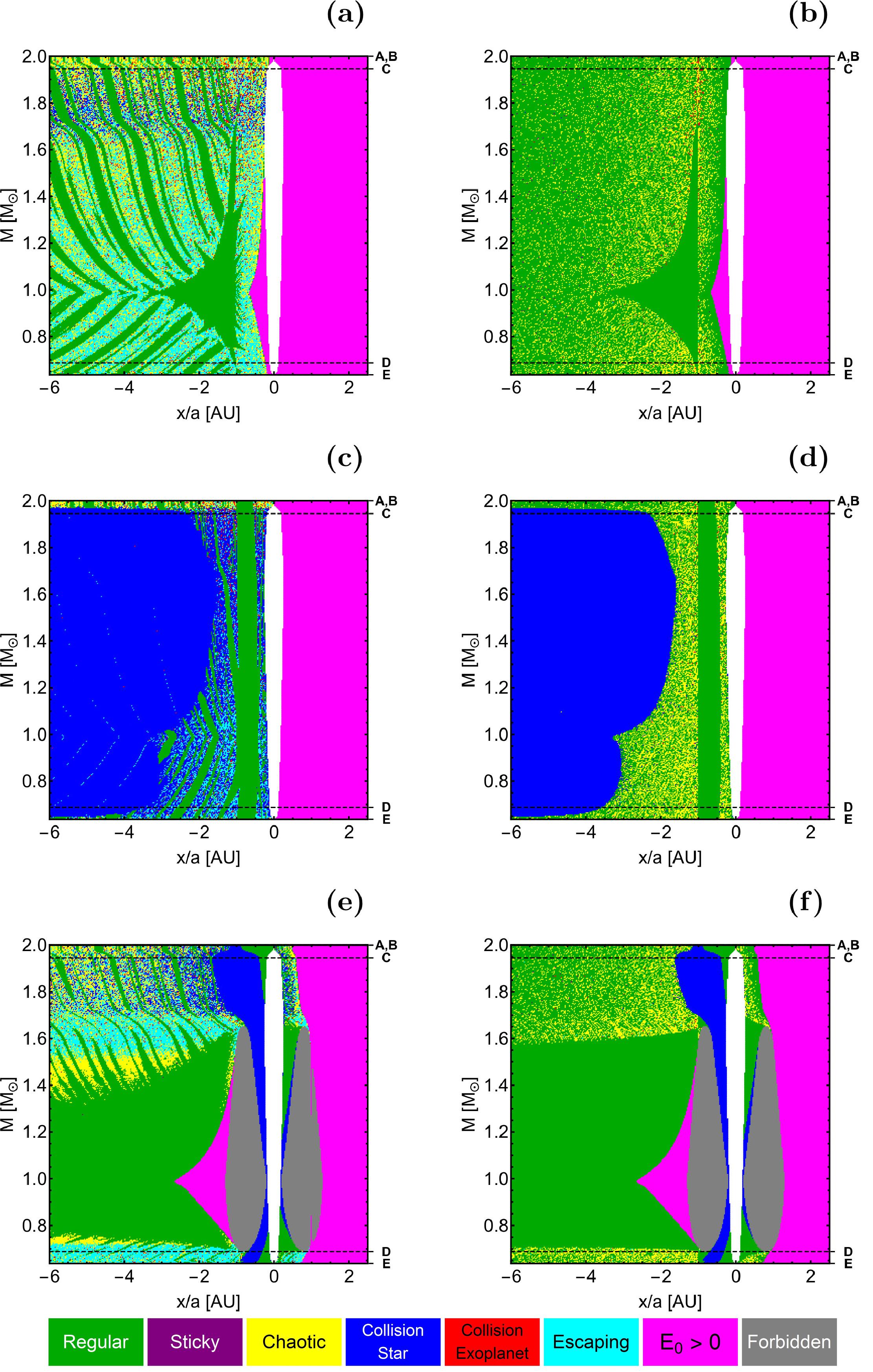}}
\caption{Same as Fig.~\ref{xm05}, but for $R_{\rm p} = 10$ mm.}
\label{xm10}
\end{figure*}

\begin{figure*}[!t]
\centering
\resizebox{0.7\hsize}{!}{\includegraphics{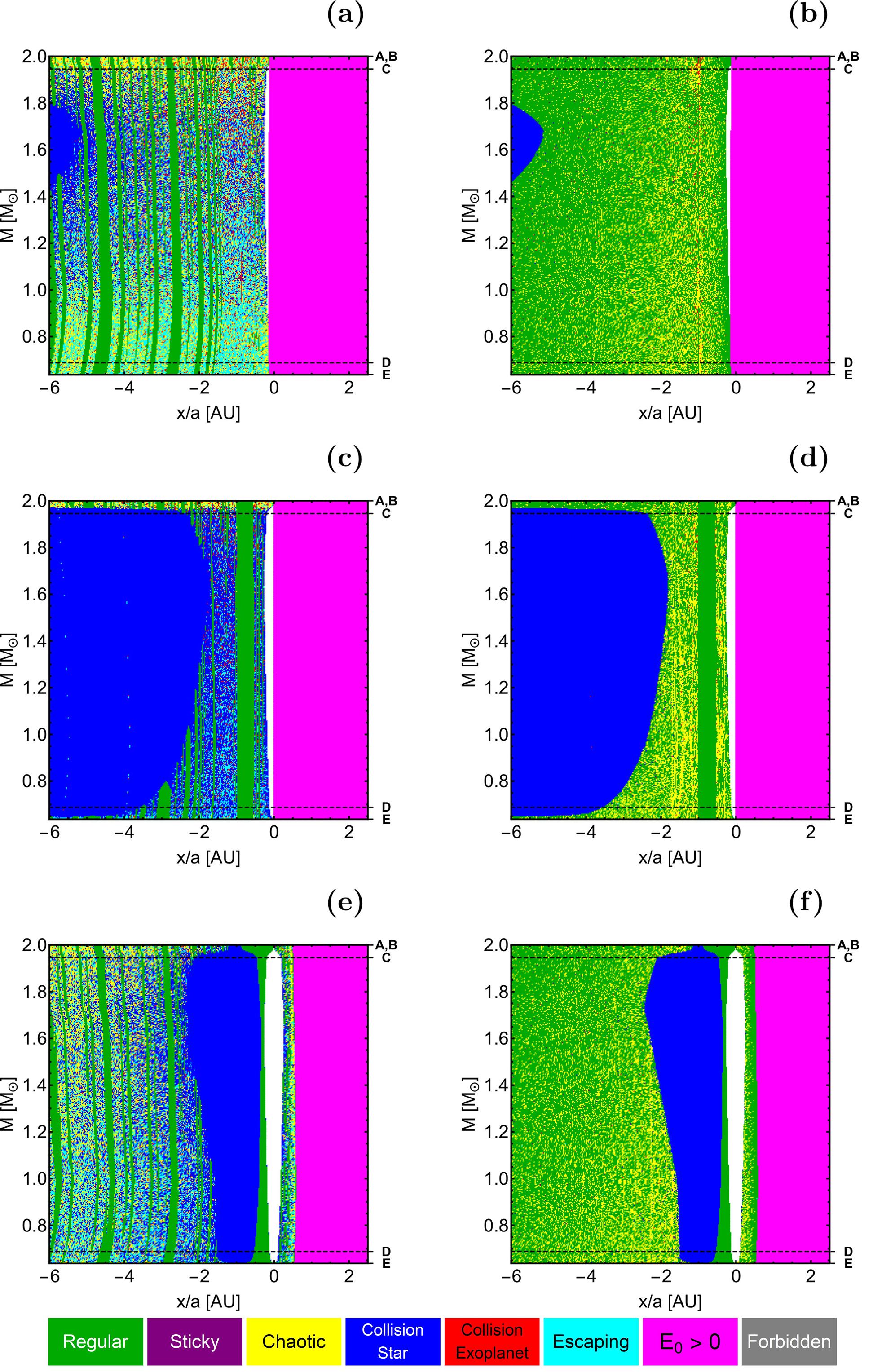}}
\caption{Same as Fig.~\ref{xm05}, but for $R_{\rm p} = 50$ mm.}
\label{xm50}
\end{figure*}

\begin{figure*}[!t]
\centering
\resizebox{0.715\hsize}{!}{\includegraphics{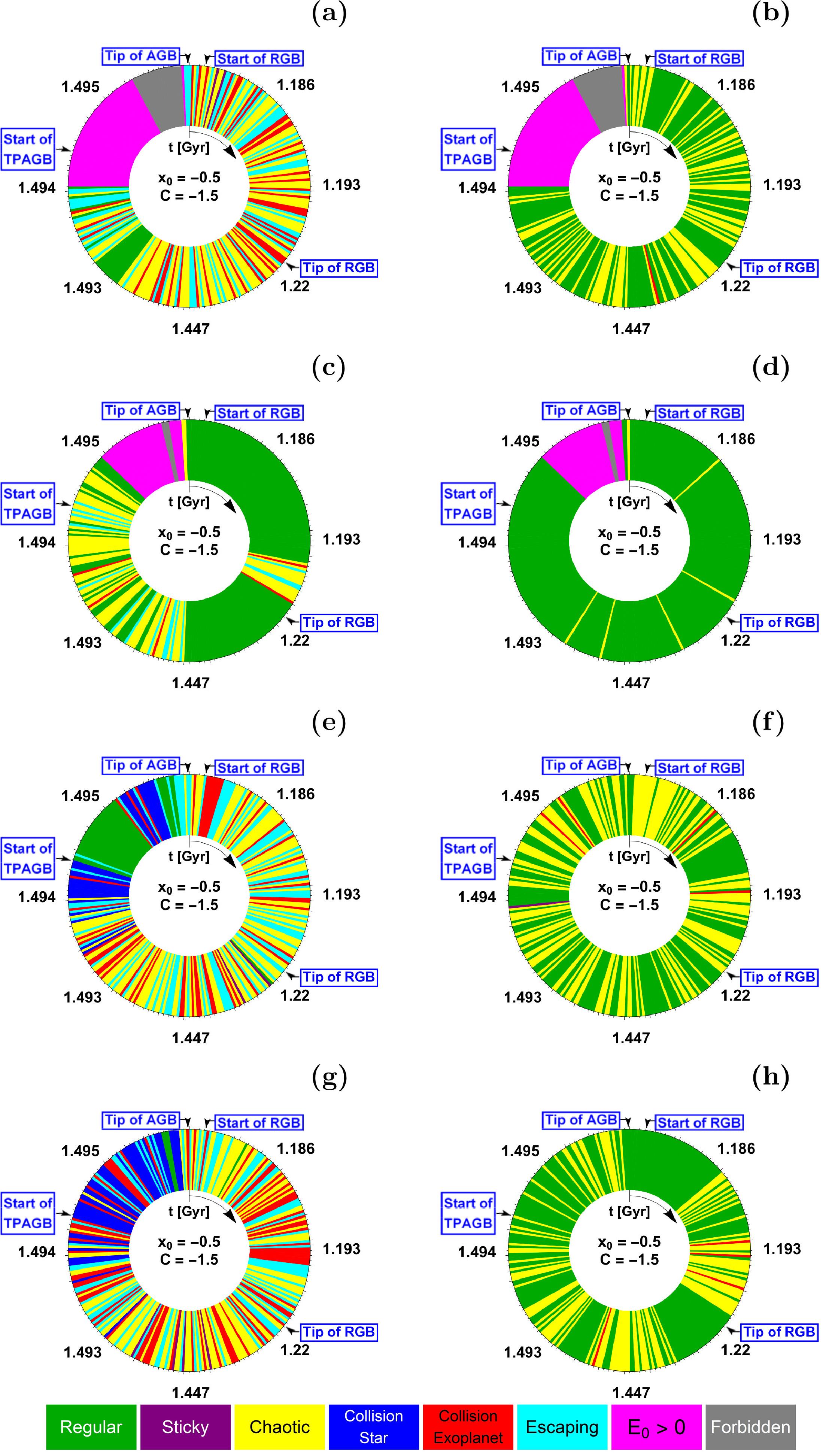}}
\caption{Orbit types as a function of time (clockwise in Gyr) for $x_0 = -0.5$ and $C = -1.5$. The simulations begin just before the start of the RGB. The left and right panels correspond respectively to Jupiter-like and Earth-like exoplanets, and the rows from top to bottom correspond to $R_{\rm p} = 1, 5, 10,$ and 50 mm.}
\label{pc1}
\end{figure*}

\begin{figure*}[!t]
\centering
\resizebox{0.714\hsize}{!}{\includegraphics{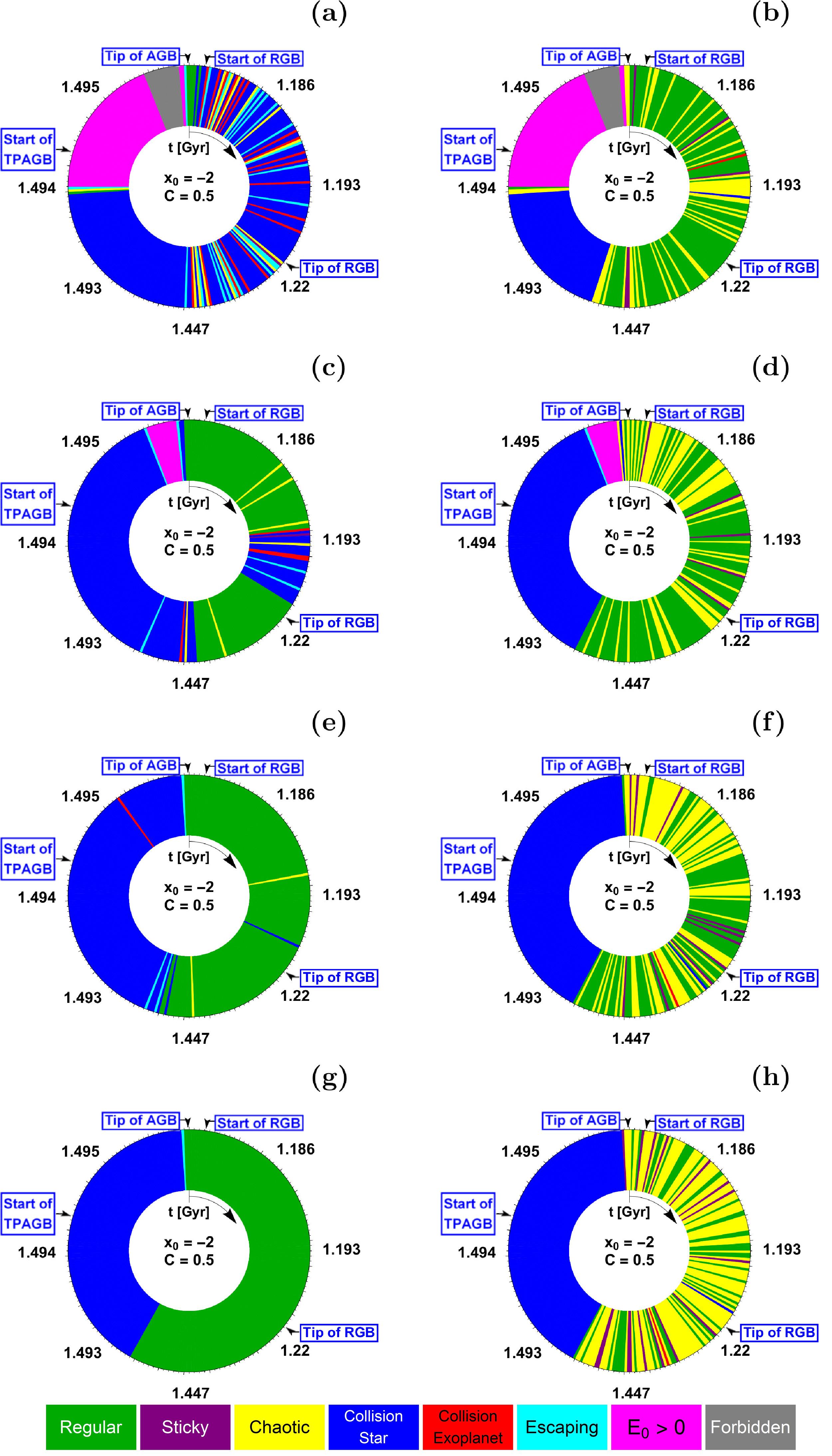}}
\caption{Same as Fig.~\ref{pc1}, but for $x_0 = -2$ and $C = 0.5$.}
\label{pc2}
\end{figure*}

\begin{figure*}[!t]
\centering
\resizebox{0.7\hsize}{!}{\includegraphics{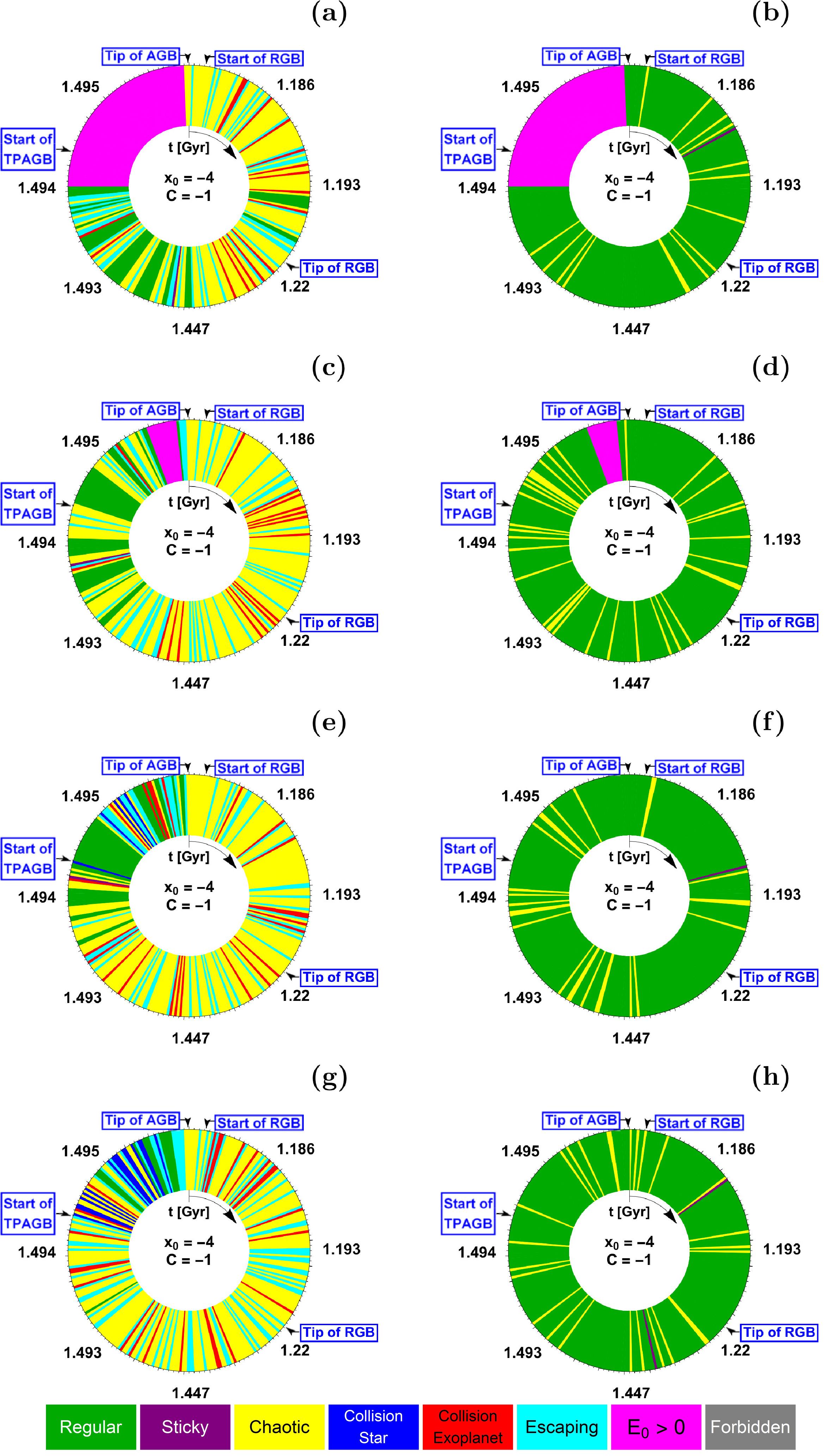}}
\caption{Same as Fig.~\ref{pc1}, but for $x_0 = -4$ and $C = -1$.}
\label{pc3}
\end{figure*}

\subsubsection{Grain properties}

The parameter of greatest interest in this study is the grain size, given by its radius $R_{\rm p}$. Also, as previously shown, the important radiation pressure factor $\beta$ is directly related to $R_{\rm p}$, as well as $L$. Further, the characteristic range of $\beta$ (which is $>0$ and not bounded from above) and the resulting implications have previously been studied for the PPRTBP \citep[e.g.,][]{zotos2015}, and hence provide a useful context for the current study.

Given the importance of the parameter $\beta$, we plot its evolution for grains of four different $R_{\rm p}$ in Fig.~\ref{evolr}. This radii range, from 1 mm to 50 mm, provides useful bounds on the maximal extent of the range of behavior one can expect, from ubiquitous escape to mostly bound and regular motion.

The figure illustrates that the highest values of $\beta$ occur just after the tip of the AGB, where the maximum luminosity is reached. The peaks of $\beta$ and the luminosity are offset by just 1600 yr, with the reason being that after the luminosity peaks, the stellar mass is still decreasing toward its eventual white dwarf value (see equation \ref{eqLM}). For context, the meaning of $\beta=1$ is the point where the gravitational and radiative forces are equal; for 1 mm grains, $\beta>0$ throughout the AGB. Importantly, at any time when $\beta > 1$, the grain would be ejected from the system.

\subsubsection{Exoplanet properties}

Because the focus of our study is on grain radius, for computational convenience we adopt two bounding values of major exoplanet mass: $1M_{\oplus}$ and $1M_{\rm Jup}$ (with corresponding radii in order to account for possible collisions). Although terrestrial planets are more likely to eventually drive white dwarf pollution \citep{frehan2014,musetal2018}, the only known major exoplanet orbiting a single white dwarf in a compact orbit is likely a giant exoplanet \citep{ganetal2019,verful2020}. Hence, both types of exoplanets are important to consider.

In all of the remaining figures (Figs.~\ref{xc01}-\ref{pc3}), the left panels all display results for $1M_{\rm Jup}$ exoplanets and the right panels all display results for $1M_{\oplus}$ exoplanets.

\subsection{Simulation results}
\label{res}

We present our results by using three types of representations in phase space: the $(x,C)$ plane, the $(x,M)$ plane, and annular charts of the final states.

\subsubsection{The $(x,C)$ plane}

The $(x,C)$-plane characterizes orbits as a function of both location and energy. We show the time evolution of these orbits in Figs.~\ref{xc01}-\ref{xc50}, where $R_{\rm p} = 1, 10, 50$ mm for, respectively, Figs.~\ref{xc01}, \ref{xc10} and \ref{xc50}. In each of these figures, the top row of panels correspond to a snapshot along the main sequence, and the next row of panels corresponds to a snapshot when $q$ is a minimum (see Table~\ref{tab1} for specific numerical values). In Fig.~\ref{xc01}, the bottom row of panels corresponds to the tip of the AGB, when $q=1$. This bottom row is {\it independent of $R_{\rm p}$}, and hence is not repeated in Figs.~\ref{xc10} and \ref{xc50}.

All three figures have common characteristics. The $(x,C)$ space is divided into several regions, where different types of motion of the test particle dominate. In particular: (i) In all cases, the right-hand side of the $(x,C)$ plane is covered by a broad uniform magenta region. Here, the corresponding initial conditions lead to immediate escape (where the initial inertial energy $E_0$ is positive); (ii) During the main sequence (top panels), in-between the planet and star (between the energetically-forbidden gray regions), the vast majority of the starting conditions lead to bound regular motion (green) around either object. In contrast, the green regions corresponding to starting conditions at $x \lesssim -2$ lead to circumbinary bound regular motion around both the planet and star; (iii) In most cases, the plots feature a claw-like structure containing a mixture of initial conditions corresponding to all possible types of motion. Discernible substructure reveals the formation of stability islands, corresponding to secondary resonant orbits. Between these stability islands of higher resonances we see an almost fractal-like mixture of chaotic, sticky, collision and escaping initial conditions \citep[see e.g.,][]{AVS01,AVS09}. Running through the middle of the claw is a uniform blue region, indicating collisional orbits with the star. Often sticky orbits do tend to ``stick'' to this blue region; (iv) In most cases, this mixture of orbits extends beyond the claw at a pinch point located at about $(x,C) = (-1,-1)$, proceeding to ``hug'' the magenta region from $-1 \lesssim x \lesssim 1$.

Now we consider the individual figures. For Fig.~\ref{xc01} along the main-sequence, the Earth-like (right panels) and Jupiter-like (left panels) cases are similar, except inside of the claw-shaped regions. Stability islands of the secondary resonances are visible only for the Jupiter-like planet, as well as the horizontal aqua string of escaping orbits. In order to explore this discrepancy further, we performed additional computations which revealed that for Earth-like planets, higher resonant regular orbits are still present even though stability islands cannot be formed. Collisional orbits with the exoplanet cover a wider swathe of phase space for a Jupiter-like exoplanet than an Earth-like exoplanet.

When the system in Fig.~\ref{xc01} has reached the time of minimum $q$, then the entire $(x,C)$ plane is covered by energetically forbidden region and basins occupied by starting conditions with $E_0 > 0$, which lead to immediate escape from the system. Hence, no grains of size $R_{\rm p} = 1$ mm can survive this violent phase.

If any grains are generated afterwards, for example from escape from the planet's surface, then the fate of these grains at the tip of the AGB is illustrated in the bottom row of Fig.~\ref{xc01}. These plots are directly comparable to the main-sequence phase (top row) of Fig.~\ref{xc01}. The two main difference are: (i) in the Jupiter-like exoplanet case (see panel (e)) the number of initial conditions which lead to escape is much higher, and (ii) for both Jupiter-like and Earth-like exoplanets, the area of the basin of collision to the star (the blue strips) are thinner. The explanation for these differences is that at the tip of AGB -- which here is defined nearly at the start of the white dwarf stage (see Fig. \ref{evols}) -- the star's radius has already reduced significantly. Consequently, the grain needs more time to reach the surface of the star and collide, and so is much more likely to escape before collision.

Now we move on to Fig.~\ref{xc10}, where the grain radius has increased by an order of magnitude to $R_{\rm p} = 10$ mm. To reiterate, the first row corresponds to the main sequence stage, while the second row to the stage where $\beta$ is maximized (during the WD stage, when $\beta=0$, the plots are the same as in the bottom row of Fig.~\ref{xc01}). For panels (a) and (b), we see that for Jupiter-like exoplanets, the $R_{\rm p} = 1$ mm and $R_{\rm p} = 10$ mm cases are very similar. However, for an Earth-like exoplanet, the structure differs.  For $R_{\rm p} = 10$ mm, along the main-sequence, the number of starting conditions leading to sticky and chaotic trajectories in the claw-shape bases substantially grows, thus reducing the number of higher resonant regular orbits.

During the time of maximum $\beta$, the differences between Figs.~\ref{xc10} and \ref{xc01} are more obvious. For the $R_{\rm p} = 10$ mm case, there exists a claw-shaped region with a thick blue strip (indicating a high number of collisional orbits with the star). For a Jupiter-like exoplanet (panel (c) of Fig.~\ref{xc10}), escaping orbits surround the strip in a highly resonant structure, whereas for an Earth-like exoplanet (panel (d) of Fig.~\ref{xc10}) nearly none of the orbits are escaping and the resonant structure is lost. Finally, both panels (c) and (d) feature white regions: these correspond to initial conditions which are inside the radius of the star or the exoplanet.

In Fig.~\ref{xc50} we present the case where the grain radius is $R_{\rm p }= 50$ mm. Along the main-sequence, the character of the orbits is similar to the $R_{\rm p} = 1$ mm and $R_{\rm p} = 10$ mm cases. However, at the epoch of maximum $\beta$, the orbital structure of the $(x,C)$ plane changes drastically. In particular, the blue strip of stellar collisional orbits is thick enough to encompass two units on the $y$-axis. Further, for a Jupiter-like exoplanet (panel (c)) the escape orbits appear to be dissected by stability islands, corresponding to secondary circumbinary resonant orbits. These stability islands vanish in panel (d), similar to the $R_{\rm p }= 10$ mm case.

\subsubsection{The $(x,M)$ plane}

The color-coded diagrams on the $(x,C)$ plane provide useful information, but are limited to particular snapshots in time. In order to overcome this limitation, we now provide similar classification grids of starting conditions, but on the $(x,M)$ plane, for given values of the Jacobi constant $C$. Thus, by using the stellar mass as a variable, we can cover almost all the evolutionary stages of the star and how they affect the final state of the test particle. We report our $(x,M)$ plane results in Figs.~\ref{xm01}--\ref{xm50}, where each successive plot corresponds to respectively $R_{\rm p} = 1, 5, 10,$ and 50 mm. The first, second and third rows of panels in each plot correspond to, respectively, $C=-1, -0.4,$ and 2.

For the smallest grain size that we consider (1 mm), Fig.~\ref{xm01} illustrates that unboundedness becomes ubiquitous soon after the star leaves the main sequence, regardless of the character of the exoplanet. The transition to escape becomes more gradual as we increase the grain radius. For $R_{\rm p} = 5$ mm (Fig.~\ref{xm05}), grains could potentially remain bound until the stellar mass decreases beyond about $1.6 M_{\odot}$. Only for $R_{\rm p} \gtrsim 10$ mm (Figs.~\ref{xm10}-\ref{xm50}) do grains appear to survive throughout stellar evolution.

Besides unboundedness, other interesting dynamical outcomes are illustrated in Figs.~\ref{xm01}-\ref{xm50}. As $C$ is increased, the number of collisional orbits with the star increases dramatically before decreasing. When the parameter space is not dominated by stellar collisional orbits, the space is instead populated by a mixture of all other types of orbits. Standout features for the $R_{\rm p} = 10$ mm and 50 mm cases are resonant structures and regions with kinks at key stellar masses in the temporal evolution of the star. All panels in Fig.~\ref{xm10} showcase kinks at the moment when $\beta$ is a maximum, with panel (a) representing a particularly striking and feature-rich example. Panels (c)-(f) of Fig.~\ref{xm50} illustrate kinks at the start of the AGB, and panels (e)-(f) of the same figure display noticeable changes in structure at the maximum $\beta$ epoch.

\subsubsection{Final state annular charts}

In addition to viewing results on the $(x,C)$ and $(x,M)$ plane, we also wish to illustrate the time evolution of the character of an orbit for a fixed $x_0$ and $C$ at high resolution. To do so we use an annular chart, where time evolves clockwise and each strip represents the orbit character at that time. We present these results in three figures (Figs.~\ref{pc1}-\ref{pc3}), each corresponding to a different value pair $(x_0,C)$ (chosen to correspond to interesting cases). Each figure contains four rows and two columns (left column for a Jupiter-like exoplanet, and right column for an Earth-like exoplanet). The four rows of panels from top to bottom respectively refer to grain radii of 1 mm, 5 mm, 10 mm and 50 mm.

In Fig.~\ref{pc1}, where $x_0 = -0.5$ and $C = -1.5$, there is a qualitative difference between Jupiter-like exoplanets (left column) and Earth-like exoplanets (right column). For Earth-like planets, the grain orbits alternate primarily between regular and chaotic motion, whereas for Jupiter-like planets, the type of orbit varies more drastically and quickly. Further, only for Jupiter-like planets are escaping motion and collisional outcomes possible.

The diagrams shown in Fig.~\ref{pc2} correspond to $x_0 = -2$ and $C = 0.5$. Here, collisional orbits with the star are prevalent during the AGB. For Jupiter-like exoplanets, throughout the RGB for $R_{\rm p} \gtrsim 5$ mm, the motion is primarily bounded and regular. For Earth-like planets, the situation is more complex, where there is a rapid change between ordered and chaotic motion; this interplay becomes more rapid with increasing grain radius.

Finally, In Fig.~\ref{pc3}, $x_0 = -4$ and $C = -1$. Here, for a Jupiter-like exoplanet, the motion of the grain is primarily chaotic, despite experiencing sudden transitions to escaping and collisional motion. In contrast, for an Earth-like exoplanet, the final state of the grain is mainly regular, with no escaping and collisional motion.

\section{Discussion}
\label{disc}

We have shown that the orbital dynamics of grains in one-planet post-main-sequence planetary systems is rich area of study and is nontrivial to characterize in parameter space. The resonant structures formed are a strong function of the mass of the exoplanet, and appear to have better definition in exo-Jovian rather than exo-terrestrial systems.

One potentially observable consequence is that planetary debris disks or rings around RGB and AGB stars \citep{bonetal2013,bonetal2014} which harbor one major planet may exhibit predictable structure, but may also be a strong function stellar age. This structure is likely to be resonant, but may also contain distinctive features which do not align with strong mean-motion commensurabilities. Particularly given the prevalence of chaotic and sticky orbits in our simulations, some features could be strongly time-varying.

Some grains which may survive for the entire main-sequence lifetime could be the result of destructive collisions in the main-sequence debris disks \citep{wyatt2008,kenbro2008,krietal2006,kraetal2013}. Further, many A-type stars with apparent ages close to their main sequence lifetimes have debris disks. Hence, some A-type stars are likely to contain significant amounts of material orbiting the central star as it leaves the main sequence \citep{suetal2006,matetal2014,hugetal2018}, as demonstrated by \cite{bonetal2013} and \cite{bonetal2014}.  The ability of grains to survive throughout the main-sequence is a function of the collisional lifetime of the disk, the fragmentation prescription adopted, and the blowout size for main-sequence luminosities and winds. Such survival is facilitated by the relatively short main-sequence lifetime of A-stars.

However, grains which are present during the giant branch phase did not necessarily survive throughout the main sequence. Also important to know is when during the giant branch phases is debris generated and initialized into the orbits on which we performed integrations. Grains could be generated easily from asteroid breakup, which is likely to be prevalent along the giant branch phases due to rotational fission from the YORP effect \citep{veretal2014c,versch2020}. The breakup could occur at any time during the giant branch phases, and is a function of the initial size distribution and location of minor planets, as well as their physical characteristics (such as initial spin and internal strength).

In fact, our results indicate that, in general, the only way for grains smaller than about 1 mm to survive until the white dwarf phase is if they are generated sufficiently late during stellar evolution to avoid escape and collisional orbits. Consequently, we speculate that contributions of submillimeter-sized particles to white dwarf pollution and debris disks are primarily second-generation: created from the destruction of minor and major planets during the white dwarf phase, and only marginally supplemented with extant grain material from the giant branch phases.

How larger grains ($R_{\rm p} \gtrsim 50$ mm) which more easily survive until the white dwarf phase contribute to observables is an open and complex question. Nevertheless, particles of 50 mm are small enough to be subject to Poynting-Robertson drag from the white dwarf luminosity, and so, depending on their separation from the white dwarf, may still contribute to both pollution and debris disk accumulation. Understanding the potential origin of grains of different sizes in these disks will become increasingly important as evidence for disk variability mounts \citep{faretal2018,xuetal2018,swaetal2019a,wanetal2019}.

Not included in our computations were the interactions between the stellar wind and the grains, an effect which has been studied in giant star planetary sytems for decades \citep[e.g.,][]{livsok1984,maretal2020}. \cite{donetal2010} estimated that the critical grain size below which entrainment is possible is a function of the density and speed of the wind as well at those of the grain. The upper size bound for entrainment is often referred to as the ``blow-out'' size; \cite{bonwya2010} showed that the blow-out size for a Sun-like star is about one micron, and is linearly dependent on stellar luminosity. Hence, at the tip of the AGB, the blowout size may exceed one mm. Gas drag on the grains is an additional effect; \cite{veretal2015c} demonstrated that gas drag is negigible for major planets, but is a stronger effect than stellar mass loss for grains.

We also assumed that the grain radii remained unchanged. In reality, the grains are sublimated to different extents depending on their material properties, seperations from the star, and stellar luminosity. Further, the reduction in surface area due to sublimation facilitates their ejection from the system. We compute the sublimation rate of the grains by using the same prescription in \cite{veretal2020} (the original version being from \citealt*{jura2008}), and find that at the tip of the AGB, a 2 g/cm$^{-3}$ grain at 5 au would be shedding at the rate of $1 \times 10^{-5}$ mm/yr; at 10 au, the rate would be reduced to $4 \times 10^{-15}$ mm/yr. The composition of the grains to some extent depends on the location of the snow line. We can approximate the maximum distance corresponding to this boundary by using the flared-disk prescription from \cite{adashu1986,kenhar1987} and \cite{chigol1997}.  Doing so yields an initial snow line at of approximately 5 au, the initial location of the planet; the subsequent evolution of the snow line outpaces the planet throughout the evolution on the giant branch.

The increased luminosity of the star also heats up the atmosphere of the gas giant planet, and the stellar wind alters the composition of its atmosphere. \cite{spimad2012} considered both these effects for Jovian planets around giant branch stars at distances of at least several au. Observationally, detecting these spectral changes may be challenging because the transit probability at such distances is so low. More detectable are planets orbiting around giant stars which will be engulfed; how tidal heating affects re-inflation mechanisms is still uncertain \citep{lopfor2016,ginsar2016,saietal2019}. Atmospheric heating also affects the prospects for habitability, both along the giant branch phases \citep{kozkal2019,kozkal2020a} and subsequent white dwarf phase \citep{fosetal2012,barhel2013,loemao2013,kozetal2018,kozkal2020b}.

Finally, we note that the dynamical architectures which allows for one major planet to perturb a smaller objects into the star are restricted \citep{bonetal2011,debetal2012,frehan2014,antver2016,antver2019} but important to consider when quantifying white dwarf pollution. The sizes of the grains also help determine which particles reach the photosphere \citep{broetal2017} and then sink into it with particular distributions and frequencies \citep{wyaetal2014,turwya2019}. In the context of our PPRTBP study, the dark blue region correspond to what would be polluting material. In particular, panels (e) and (f) of Fig.~\ref{xc01} provide a glimpse of the locations and energies of grains which may represent polluters just after the star has become a white dwarf.

\section{Conclusion}
\label{conc}

Understanding the evolution of debris in post-main-sequence planetary systems will help us interpret the mounting observations. Here, we have performed a dynamical analysis of grains in giant branch systems containing one major planet which survives stellar evolution; this situation is reflective of the majority of currently-observed giant branch planetary systems.

By adopting the Photogravitational Planar Restricted Three-Body Problem (PPRTBP) as our tool, we were able to explore an extensive range of parameter space and characterize grain orbits as bound, unbound, collisional, chaotic and sticky. Overall, we identified the grain radii size range of 1 mm -- 5 cm as the boundary demarcating where grains are predominantly radiatively blown and gravitationally scattered away. Consequently, in one-planet white dwarf systems, we speculate that contributions from first-generation submillimeter-sized grains are negligible.

Our analysis strongly suggests that the type of exoplanet (Jupiter- or Earth-like) strongly influences the motion of the grain. In particular, for the case of an Earth-like exoplanet we found the following differences with respect to the case with a Jupiter-like exoplanet: (i) the stability islands corresponding to higher resonant orbits appear well-formed only for Jupiter-like exoplanets, (ii) in the case of an Earth-like exoplanet, the amount of escaping and collision to secondary trajectories is significantly reduced, (iii) when an Earth-like exoplanet is present, the color-coded maps appear more chaotic, and (iv) the variation of the final state of the grain, during the evolution of the star, is much more violent in the case of a Jupiter-like exoplanet; in the case of an Earth-like secondary, the final state of the grain mainly changes between regular and chaotic bounded motion.

\section*{Acknowledgments}
We thank the referee for a very helpful report, which has significantly improved the manuscript. The second author (DV) gratefully acknowledges the support of the STFC via an Ernest Rutherford Fellowship (grant ST/P003850/1).


\end{document}